\documentclass[twocolumn]{aastex631}

\received{Augusts 5, 2022}
\revised{November 16, 2022}
\accepted{\today}

\shorttitle{Locating Exoplanets Using Machine Learning}
\shortauthors{Terry et al.}

\begin{document}



\title{Locating Hidden Exoplanets in ALMA Data Using Machine Learning}

\author[0000-0002-8590-7271]{J. P. Terry}
\affil{Department of Physics and Astronomy, The University of Georgia, Athens, GA 30602, USA.}
\affil{Center for Simulational Physics, The University of Georgia, Athens, GA 30602, USA.}

\author[0000-0002-8138-0425]{C. Hall}
\affil{Department of Physics and Astronomy, The University of Georgia, Athens, GA 30602, USA.} 
\affil{Center for Simulational Physics, The University of Georgia, Athens, GA 30602, USA.}

\author[0000-0002-9426-3789]{S. Abreau}
\affil{Department of Medicine, Division of Cardiology, University of California, San Francisco, CA, 94143, USA.}
\affil{Cardiovascular Research Institute, San Francisco, CA, 94158, USA.}

\author[0000-0002-6222-8102]{S. Gleyzer}
\affil{Department of Physics and Astronomy, The University of Alabama, Tuscaloosa, AL 35487, USA}



\begin{abstract}

Exoplanets in protoplanetary disks cause localized deviations from Keplerian velocity in channel maps of molecular line emission. Current methods of characterizing these deviations are time consuming, and there is no unified standard approach. We demonstrate that machine learning can quickly and accurately detect the presence of planets. We train our model on synthetic images generated from simulations and apply it to real observations to identify forming planets in real systems. Machine learning methods, based on computer vision, are not only capable of correctly identifying the presence of one or more planets, but they can also correctly constrain the location of those planets.

\end{abstract}




\section{Introduction} \label{sec:intro}
In recent years, the kinematic analysis of protoplanetary disks has proven important in demonstrating that ring and gap-like features are indeed due to the presence of unseen planets~\citep{Pinte2018, Teague2018}. The spiral wake from planets causes localized deviations from Keplerian motion known as ``kinks''. Other mechanisms, such as gravitational instability \citep{hall2020}, vertical shear instability~\citep{Barraza2021}, and magnetorotational instability~\citep{Simon_2015}, may leave similar kinematic imprints.

\par 
Recent kinematic observations of the HD 97048 system revealed a localized velocity deviation signaling the presence of a forming exoplanet~\citep{Pinte2019} too faint to be observed directly at $\sim$mm wavelengths. To estimate the location of the forming exoplanet, several computationally expensive hydrodynamics simulations were run, synthetic observations were generated and compared ``by-eye'' to real observations to obtain the best fit. 

\par 

\par 

Machine learning offers an alternative route. The use of machine learning in astronomy has exploded in the last decade~\citep{jo2019, moller20, alexander_2020}. The study of protoplanetary disks largely relies on images, so machine learning techniques for computer vision are perfectly suited for application to these kinematics observations. Computer vision is a field of AI that focuses on image and video analysis. From simple transcription of hand-written digits~\citep{mnist} to locating and classifying objects pixel-by-pixel~\citep{segmentation}, computer vision has a wide range of uses. The introduction of advanced architectures, such as residual~\citep{resnet} and inception~\citep{inception} blocks, along with qualitatively new algorithms, such as the vision transformer~\citep{Vaswani_2017, Dosovitskiy_2020}, has led to computer vision models exceeding human experts at times~\citep{Zhou2021}.

\par 
Given the applicability and strong performance of computer vision techniques, we have applied them to the kinematic analysis of protoplanetary disks. \par 
Previous work has successfully applied deep learning to the problem of inferring properties of planets within protoplanetary disks~\citep{Auddy2021, Auddy2022} using models trained on simulated data, but these efforts do not use kinematic information and instead rely on gaps as the signatures. We demonstrate that a trained network can detect the presence of planets in large datasets much faster than a human, recover all previously identified planets, and possibly identify ones that have been missed in previous analyses. Inferring other properties, such as the number of planets and their masses, may be within the reach of machine learning techniques as well, which we leave to future work. This paper presents a proof of concept application of a relatively simple machine algorithm applied to protoplanetary accretion disks. 



\par 
The paper is arranged as follows: Section~\ref{sec:methods} describes the training data and the details of the machine learning models and their training. Section~\ref{sec:results} presents the results, interpretations, suggestions for future work, and limitations. Section~\ref{sec:conclusion} gives our conclusions.

\section{Methods} \label{sec:methods}


Training a model to predict the presence of a planet requires a large amount of data spanning a large parameter space. We run 1,000 simulations in this work. The parameter space is sampled using a Latin Hypercube (LHC)~\citep{LHC}, which accounts for previously sampled points to evenly distribute values across the entire parameter space. We sample values from observed ranges inferred from disk surveys~\citep{Andrews2009, DSHARP,DSHARP_2} and widely accepted simulational parameters~\citep{Pinte2018, Pinte2019, Pinte2020}.

\par
Table~\ref{tab:lhc} shows the variables sampled and their minimum and maximum value. Approximately one third of the simulations have no planets in to address any class imbalance (i.e., not having an equal number of ``at least one planet" class and ``no planet" class simulations) while ensuring diversity in the number of planets. After sampling the stellar mass for use in the simulation, stellar temperatures and radii, both of which are used in the radiative transfer calculations, are determined using isochrones calculated by~\citet{siess}. 
\begin{table*}

    \centering
    \begin{tabular*}{0.92\linewidth}{|l|l|l|l|l|}

    \hline 
    Parameter       &   Variable &   Minimum Value      &     Maximum Value & Use \\
    \hline 
    Number of planets  &  $N$ &  0    &    4  &  \texttt{PHANTOM}, \texttt{MCFOST} \\
    Disk Inner Radius (au) & $R_{out}$  &   150    &  250 &  \texttt{PHANTOM}, \texttt{MCFOST}\\
    Disk Outer Radius & $R_{in}$ & $0.1R_{out}$ & $0.1R_{out}$ &\texttt{PHANTOM}, \texttt{MCFOST}\\
    Planetary Mass (M$_{\oplus}$) & $M_{planet}$  & 5  & 120 & \texttt{PHANTOM}, \texttt{MCFOST}\\
    Stellar Mass (M$_{\odot}$) & $M_{\star}$   &  0.5  & 2.0 &  \texttt{PHANTOM}, \texttt{MCFOST}\\
    Planet Semimajor Axis (au) & $a$  &   $R_{in} + 0.2R_{out}$  & $R_{in} + 0.8R_{out}$ & \texttt{PHANTOM}\\
    Mass Ratio & M$_{\rm{disc}}$ / $M_{\star}$  & 5$\times 10^{-4}$   &  $10^{-2}$ & \texttt{PHANTOM}\\
    Disk Viscosity & $\alpha_{\rm{SS}}$  & 5$\times 10^{-4}$   &  2.5$\times 10^{-3}$ & \texttt{PHANTOM}\\
    Surface Density Profile ($\Sigma = \Sigma_{0} R^{-p}$) & $p$   &  0.1   &   1.0& \texttt{PHANTOM}\\
    Sound Speed Profile ($c_{s} = c_{s, 0} R^{-q}$)& $q$     &   0.1   &  0.75& \texttt{PHANTOM} \\
    Stellar Age (Myr) &  $t$  &  0.5    &  5.0  & \texttt{MCFOST}\\
    Inclination (degrees) & $i$ &  10    &  80 & \texttt{MCFOST} \\
    Azimuth (degrees)     & $\phi$ & 0     &   360 & \texttt{MCFOST}\\
    Distance (pc)      &  $d$ &  100    &   200 & \texttt{MCFOST} \\
    Spectral Resolution (km/s) & $\Delta v$ &   0.01   &  0.1 & Synthetic Observations \\
    Spatial Resolution (arcseconds) & \arcsec &   0.025   &   0.25 & Synthetic Observations\\
    RMS Noise (\%)  &  $\delta$ &   0.05     &    20 & Synthetic Observations \\
    \hline
    \end{tabular*}
    
\caption{Simulation parameter ranges for the LHC sampling and the stage at which they were applied.}
\label{tab:lhc}
\end{table*}
%


\subsection{Hydrodynamical simulations} \label{ssec:sims}


\par
We run 1,000 3D smoothed particle hydrodynamics (SPH) simulations using the \texttt{PHANTOM}~\citep{phantom18} code. System properties, such as stellar mass, disk properties, observational resolution etc. were LHC-sampled from Table \ref{tab:lhc}. A significant fraction (25\%) of these simulations were withheld for testing. Each simulation uses $10^{6}$ SPH particles and evolves for 100 orbits. A given planet's accretion radius is set to 1/4 of its Hill radius. The surface density profile is set to $\Sigma \propto R^{-p}$, and the sound speed is set to $c_{\rm{s}}\propto R^{-q}$.


\subsection{Velocity channel maps} \label{ssec:vel}

The simulation results are used to create velocity channel maps in only $^{13}$CO to keep the size of the parameter space manageable, although $^{12}$CO and C$^{18}$O are also reasonable choices. 
The channel maps are created using the \texttt{MCFOST} radiative transfer code~\citep{Pinte2006, Pinte2009}. For each simulation, multiple points between 10 and 100 orbits are chosen randomly. Since one third of the simulations have no planet, six snapshots are chosen from each of those simulations while only two snapshots are chosen for simulations with planets. This eliminates the class imbalance. From each snapshot, a velocity channel cube is made using either the $^{13}$CO J 2$\rightarrow$1 (220.4 GHz) or 3$\rightarrow$2 (330.6 GHz) transition, which is randomly sampled. These transitions are used because they were also used to identify the planets in HD 163296 and HD 97048~\citep{Pinte2018, Pinte2019}. The final result is a velocity cube, where each velocity channel is an image with 600x600 pixels at a resolution of 1 au/pixel. This velocity cube is used as the model input. Channel maps are calculated using $10^8$ photon packets. The dust-to-gas ratio is 1:100, and dust is composed of carbon and silicates~\citep{Draine1984}. Stellar radii and temperatures are calculated with isochrones given by~\citet{siess} using the LHC-sampled masses and ages. A spherical beam is used.

\par

\subsection{Simulated Observations} \label{ssec:simulated_obs}

The velocity channel maps are convolved spatially and spectrally, and noise is added by LHC sampling from Table~\ref{tab:lhc}. RMS noise is added independently for each pixel, which more faithfully replicates observational effects such as hot pixels. This method results in an average RMS of $\sim 10\%$ for each image.

\begin{figure*}[t!]
    \centering
    \includegraphics[width=0.95\linewidth]{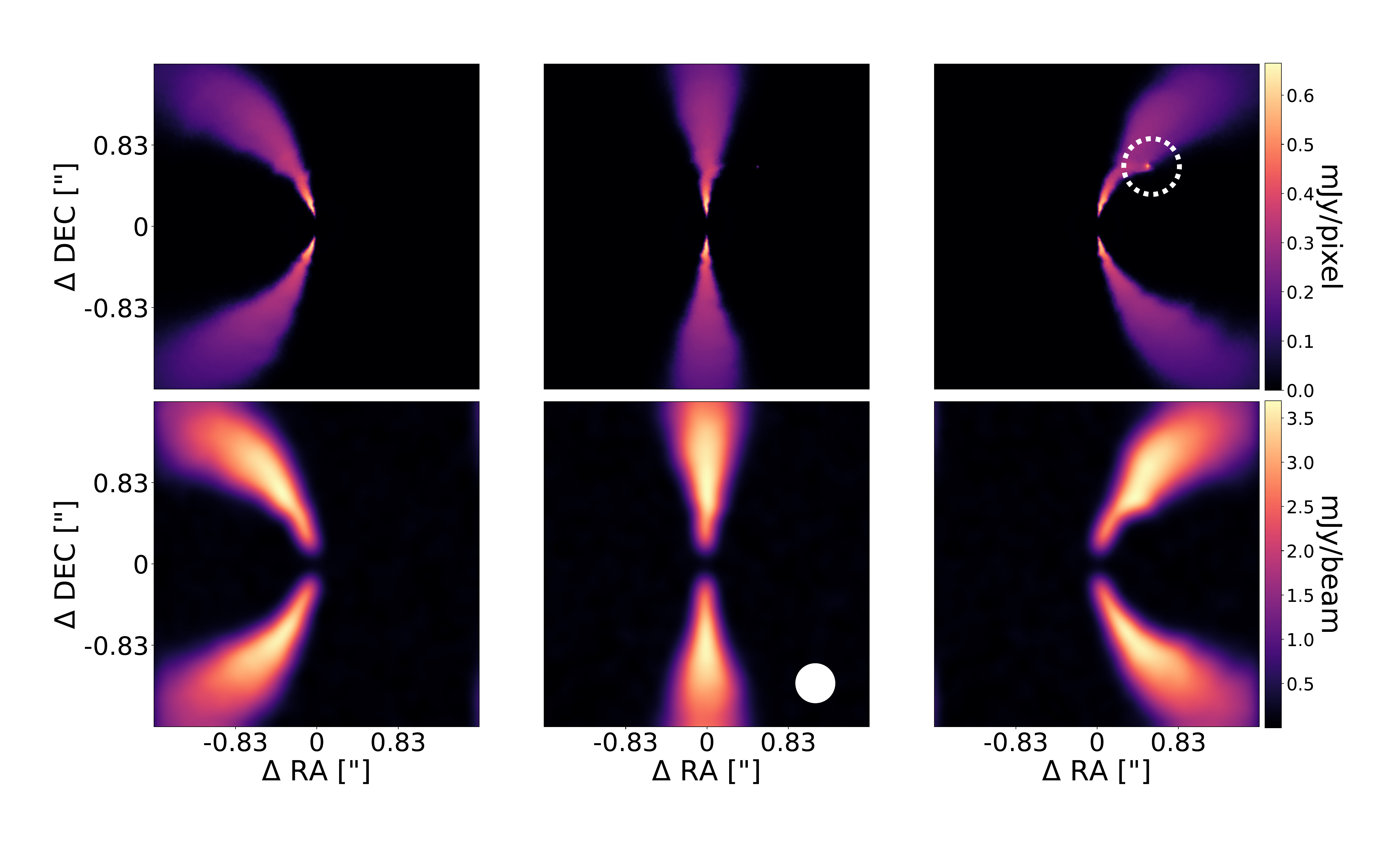}
    \caption{Example raw (top row) and convolved, noisy (bottom row) channel maps in a disk with a planet present. The planet (circled in white) is visible as a kink in the right column. The opposite velocity channel is shown in the left column, and the systemic channel is shown in the middle column. The beam size is indicated in the bottom middle image. This disk is one of the smallest and farthest simulated and is observed with some of the worst spatial resolutions, which is why the beam is so large.}
    \label{fig:planet_channels}
\end{figure*}
\begin{figure*}[t!]
    \centering
    \includegraphics[width=0.95\linewidth]{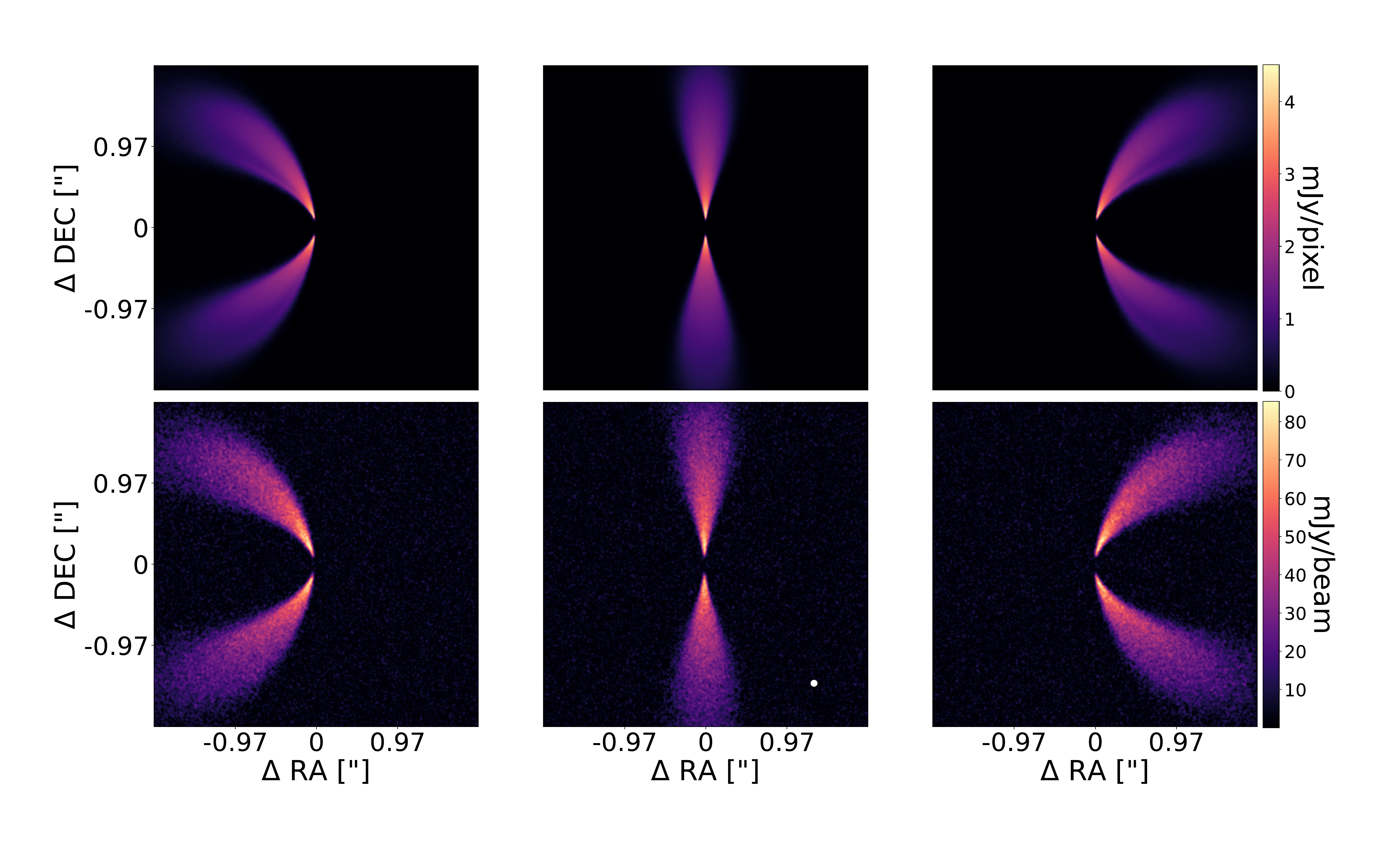}
    \caption{Example raw (top row) and convolved, noisy (bottom row) channel maps in a disk without a planet present. The beam size is indicated in the bottom middle image.}
    \label{fig:no_planet_channels}
\end{figure*}

Except for the specific values, this process largely follows that done in~\citet{terry2022}. Figures~\ref{fig:planet_channels} and~\ref{fig:no_planet_channels} show the results of convolving selected velocity channels for a disk with and without a planet, respectively. Figure~\ref{fig:planet_channels} shows that while the perturbation is strongest in the velocity channel that covers the planet (rightmost column), the perturbation is not necessarily completely localized and can be visible in distant regions of the disk. This is an expected behavior. The deviation is strongest at the location of the planet, but the spiral wake leaves further imprints as it disperses throughout the disc.
%
%
\par
By itself, a single velocity channel is poorly suited for our method. Only a few channels display kinks, and channels with kinks can neighbor the exact channel containing the planet. We therefore stack all C velocity channels into a single image of dimension (H, W, C), where H and W are the height and width of the image (600x600 pixels for these purposes). 
\par
The number of velocity channels in an observation varies, but mostly fall in the range $\sim 40-80$ channels that cover the disk ~\citep{DSHARP, MAPS}. We therefore use C=47, 61, and 75 channels for each observation. The number of channels must be odd to cover the disk symmetrically and still retain the systemic ($\delta v = 0$) channel. 
The input velocity of the simulated channels ranges from $|\delta v| = |v_{systemic} - v| \leq 2.5$ km/s to fully cover the disk.

\subsection{Machine Learning Models}\label{ssec:models}

The input for all models is a (600x600xC) image, and their output is a two-component vector with softmax activation~\citep{softmax}. The softmax function maps the output vector such that the sum of the components equals one. Each component can be interpreted as the predicted probability that the input belongs to the corresponding class (i.e., ``contains at least one planet'' class or ``does not contain a planet'' class). An Adam optimizer~\citep{adam} is used to add momentum, which remembers previous gradients, and gradient scaling to a simple gradient descent algorithm. This helps escape local minima and improves the speed and range of convergence. The cross-entropy loss~\citep{cross_entropy} penalizes confident predictions that are incorrect. Training stops when there is no improvement in the validation loss after eight complete epochs, i.e. eight sweeps through the training data. 

\par 
Multiple architectures are used to compare performance on two tasks: predicting if the system hosts at least one planet and determining its azimuthal location.

\par
We use two different models based on PyTorch~\citep{pytorch} implementations: EfficientNetV2~\citep{effnet_v2} and RegNet~\citep{regnet}. Neither model uses the default hyperparameters or pre-trained weights. 

\par
EfficientNetV2 is an updated version of EfficientNet~\citep{effnet} that maximizes performance and minimizes the number of trainable parameters and training time. RegNet, on the other hand, seeks to maximize performance by adding variants of convolutional recurrent neural networks to ResNet~\citep{resnet}. This can extract greater spatial information at the cost of significantly more trainable parameters.

\par
We perform Bayesian hyperparameter tuning using \texttt{WANDB}~\citep{wandb} to find separate sets of hyperparameters that minimize the validation loss for C=47, 61, and 75. The default parameters for these sweeps are based 
on default versions ``EfficientNetV2 S'' and ``RegNetY 16GF.''

\par
Each architecture is evaluated for accuracy and area under the receiver operating characteristic curve (AUC)~\citep{auc}. Accuracy in this context is defined as whether a model correctly predicted that a system with planet(s) has at least one planet, and similarly for systems with no planets. We choose AUC because it gives a robust measure of the performance of the model across different decision thresholds, i.e. the softmax threshold required to claim the presence of a planet. Substantial confidence is required to claim a planet detection, but setting the decision threshold too high can lead to false negatives. False positives may be more acceptable because further analysis can overrule them, whereas a false negative may lead to no further analysis and result in the planet not being detected. The optimum threshold for this boundary is open for discussion by the community. We show multiple thresholds (50\%, 75\%, 90\% and 95\%) in Figures~\ref{fig:conf_mat} and~\ref{fig:metrics} to demonstrate how the threshold for claiming the presence of at least one planet affects the results .


\par
While some protoplanetary disks exhibit gaps in continuum that are highly suggestive of a planet, this does not help one find the azimuthal location. Planet-induced non-Keplerian kinks tend to be strongest near the planet, so our model must determine if and where the kink is in our data in order to give information on the azimuthal location.
\par
One simple approach to this is to look at the activations inside the model itself. The activation strengths highlight regions which the model finds particularly informative for classification. EfficientNetV2 uses SiLU activations~\citep{silu}, and RegNet uses ReLU~\citep{relu}.%
%
We use mean-subtracted activations to achieve this.
For a given pixel with an activation value, $x$, the mean-subtracted activation, $x'$, is defined as $x' = |x - \langle x \rangle |$, where $\langle x \rangle$ is the mean activation over the entire image. Activations occur most strongly at the location of the strongest deviation from Keplerianity, i.e. the kink.

\section{Results and Discussion} \label{sec:results}


\subsection{Machine learning Models} \label{ssec:ml_results}

\begin{table*}

    \centering
    
    \begin{tabular*}{0.79\linewidth}{lcccc}

    \hline 
    Model  &     Parameters & Accuracy (50\%)   &    Accuracy (95\%) & AUC  \\
           &      (millions)          &   ($\%$)   & ($\%$)   &    \\
    \hline 
    
    EfficientNetV2: 47 Channels & 20.2   &   97 $\pm$ 0.5 & 96 $\pm$ 0.5 & 0.99 $\pm$ 0.002     \\
    EfficientNetV2: 61 Channels & 20.2   &   97 $\pm$ 0.5 & 94 $\pm$ 0.7 & 0.99 $\pm$ 0.003   \\
    EfficientNetV2: 75 Channels & 20.2   &   93 $\pm$ 0.7 & 88 $\pm$ 0.9 & 0.98 $\pm$ 0.003    \\
    RegNet: 47 Channels         & 51.0   &   78 $\pm$ 1.1 & 65 $\pm$ 1.3 & 0.86 $\pm$ 0.010    \\
    RegNet: 61 Channels         & 62.8   &   98 $\pm$ 0.4 & 96 $\pm$ 0.6 & $>0.99$ $\pm$ 0.001 \\
    RegNet: 75 Channels         & 114    &   95 $\pm$ 0.6 & 92 $\pm$ 0.7 & 0.98    $\pm$ 0.003 \\
    
    \hline
    \end{tabular*}
    
\caption{Final model descriptions and metrics. 95\% confidence intervals are calculated by bootstrapping each metric 1,000 times using random selections of 80\% of the test data. The percentages next to the accuracy labels denote the decision threshold.}
\label{tab:hparams}
\end{table*}

We train six models in total: an EfficientNetV2 and a RegNet for 47, 61, and 75 velocity channels. 
Using a typical decision threshold of 0.5, each model, other than the 47-channel RegNet, has an accuracy of at least 93\%. When this decision threshold is increased to 0.95, four models have an accuracy of at least 92\%. Five models have an AUC of at least 0.98. This confirms that our models have learned to correctly detect the presence of planets using kinematic signatures. 

\par
Table~\ref{tab:hparams} shows the number of parameters as well as relevant performance measures for each final model. All EfficientNetV2 models converge to a similar number of trainable parameters, whereas the number of parameters for the RegNets increases sharply with the number of channels. RegNet is a larger and more complicated architecture, so this is expected. The performance of the RegNet models varies widely compared to the EfficientNetV2 models. The accuracy of EfficientNetV2 models varies by at most 8\%, whereas RegNet varies by as much as 31\%. By all performance metrics, while the average EfficientNetV2 outperforms the average RegNet, the 61-channel RegNet outperforms all other models. 

\begin{figure*}
    \centering
    \includegraphics[width=1.0\linewidth]{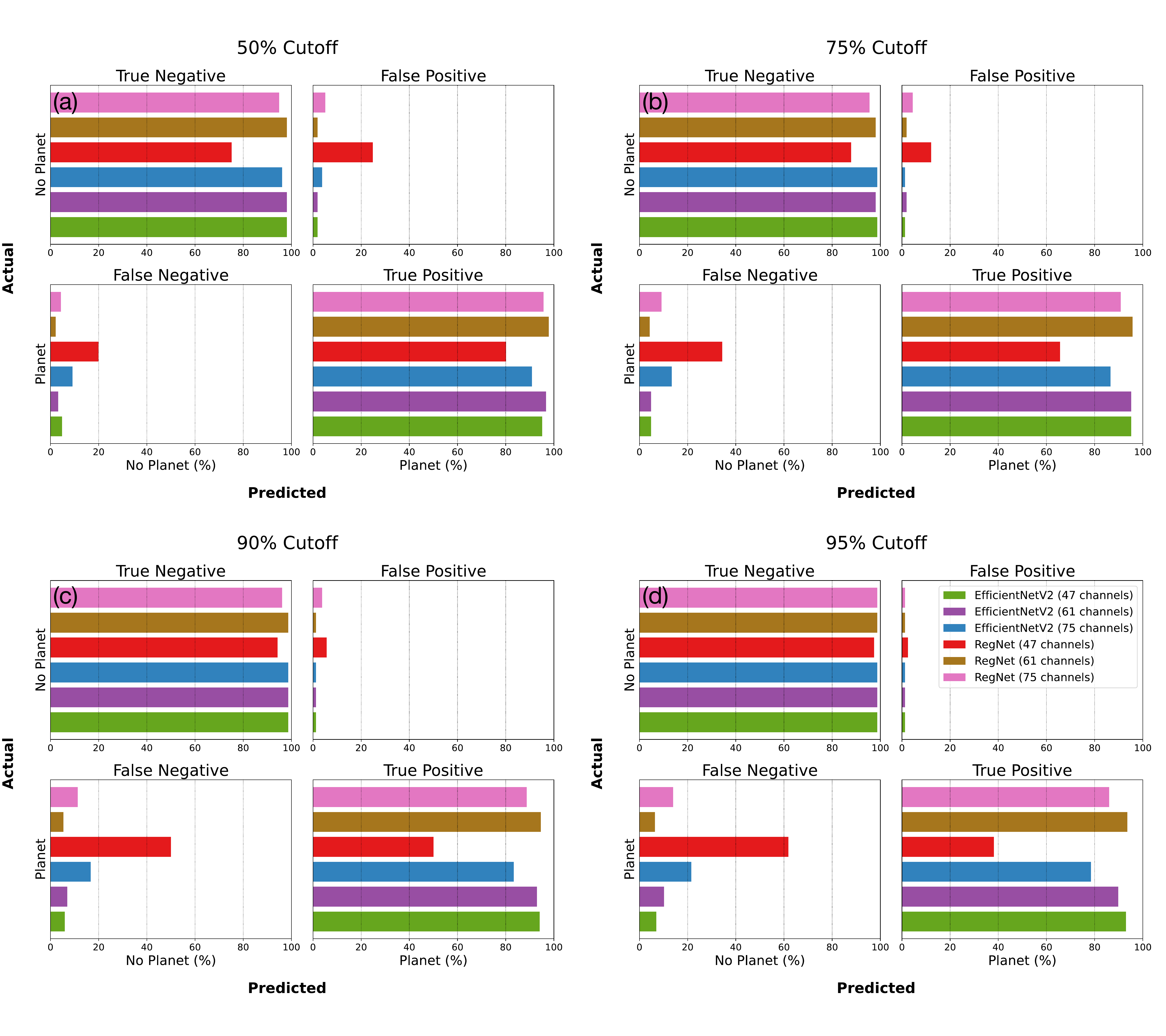}
    \caption{Confusion matrix from withheld test set for all models. The $x$ axis is the percent of disks in that class that were predicted accordingly. (a): Using a 50\% decision threshold. (b): Using a 75\% decision threshold. (c): Using a 90\% decision threshold. (d): Using a 95\% decision threshold. The top left and bottom right blocks of each figure show the counts of true negatives and true positives, respectively. The top right and bottom left blocks of a figure show the false positives and false negatives, respectively. A perfectly accurate model will have no entries in the top right and bottom left.}
    \label{fig:conf_mat}
\end{figure*}

\begin{figure*}
    \centering
    \includegraphics[width=1.0\linewidth]{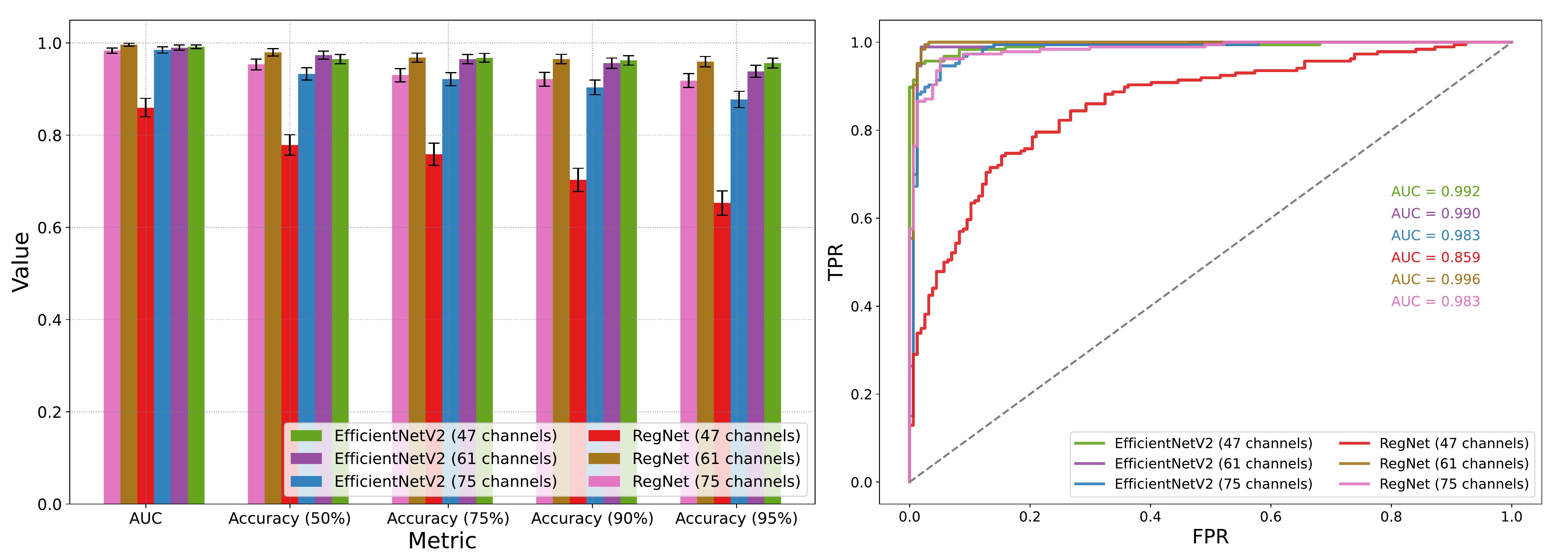}
    \caption{Left: Various metrics calculated from the withheld test set for all models. Right: Corresponding ROC curves. Error bars show 95\% confidence intervals that are calculated by bootstrapping each metric 1,000 times using random selections of 80\% of the test data. The percentages next to the accuracy labels denote the decision threshold.}
    \label{fig:metrics}
\end{figure*}

\par
Figure~\ref{fig:conf_mat} shows the confusion matrix, which describes the distribution of correct and incorrect predictions, for all models applied to the withheld test set. Panel (a) shows the results using a 50\% decision threshold, i.e. a planet is predicted if the confidence is over 50\%. Panels (b), (c), and (d) show the same for cutoffs of 75\%, 90\%, and 95\%. Figure~\ref{fig:metrics} gives several relevant metrics for model performance and all ROC curves. Claiming a planet is present should come with high confidence, so a 50\% decision threshold isn't necessarily what should be used in practice. It is encouraging that, for all other models than the 47-channel RegNet, there are no qualitative changes in the results when a higher decision threshold is used; the main effect is simply increasing the number of false negatives and decreasing the number of false positives. 

\par
\begin{figure*}
    \centering
    \includegraphics[width=1.0\linewidth]{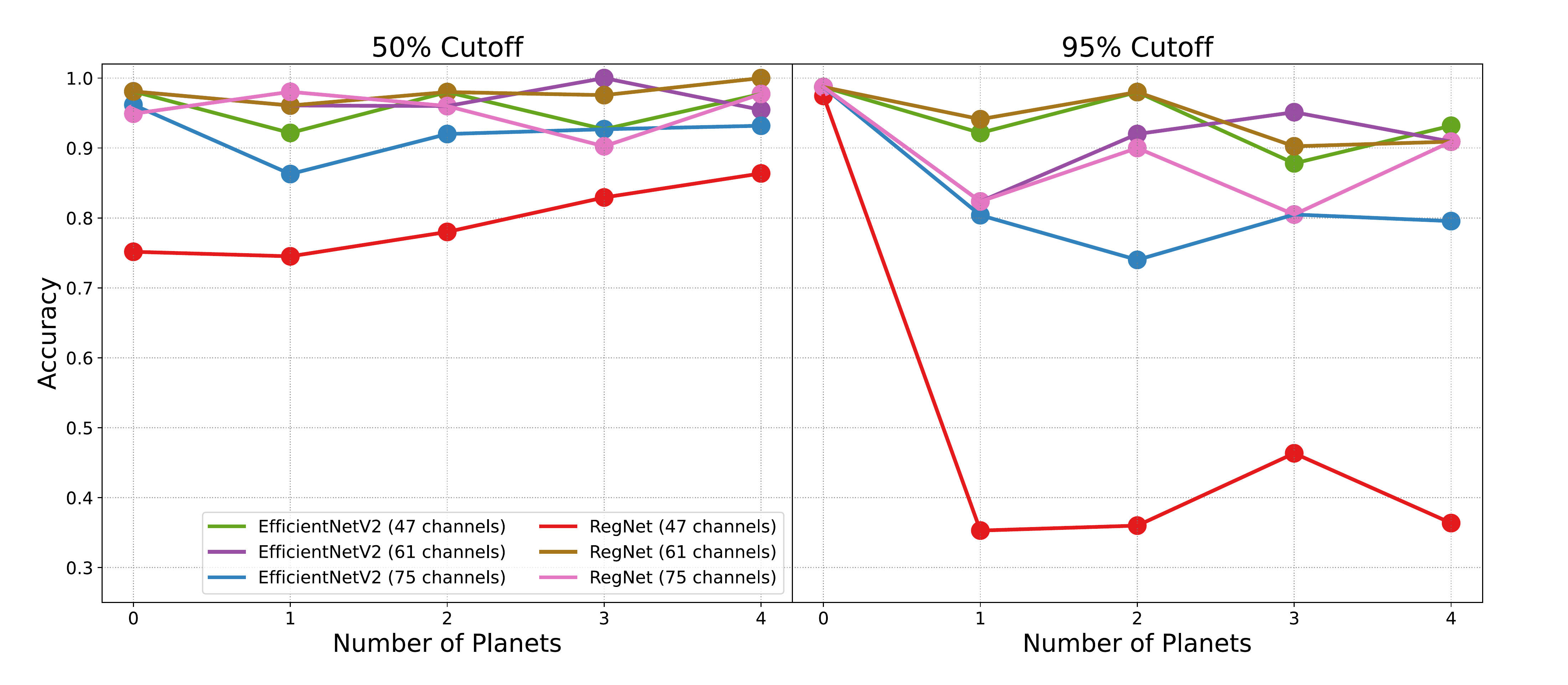}
    \caption{Accuracy of all models for a given number of planets. Left: 50$\%$ decision threshold. Right: 95$\%$ decision threshold}
    \label{fig:acc_N}
\end{figure*}

\begin{figure}
    \centering
    \includegraphics[width=1.0\columnwidth]{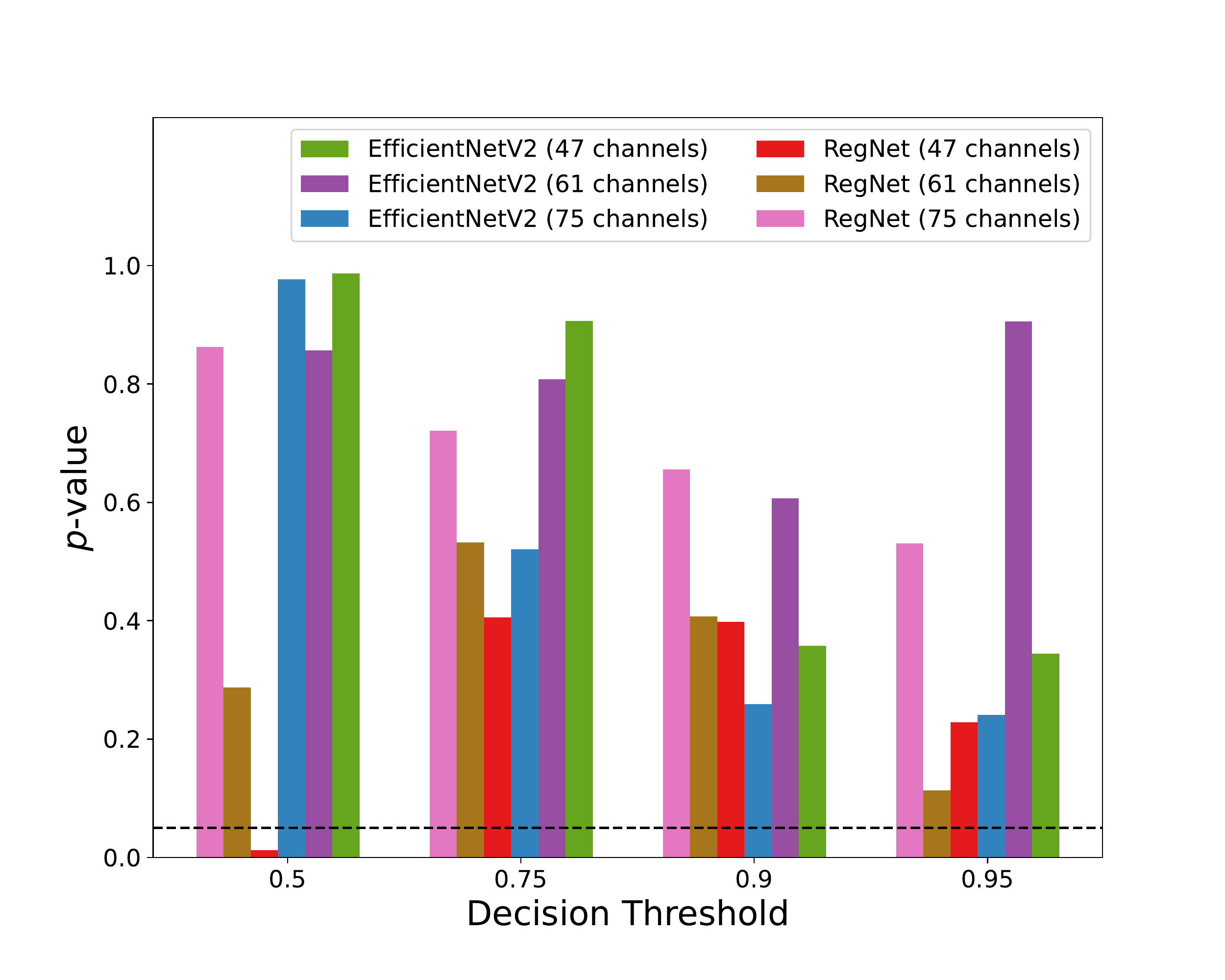}
    \caption{$p$-values from testing the correlation between accuracy and the number of planets at different decision thresholds. The black line is $p=0.05$, which is the standard cutoff for statistical significance.}
    \label{fig:p_values}
\end{figure}

 
If there are multiple planets in a disk, then multiple kinks may be present. It is therefore prudent to check that the quality of the predictions does not depend on the number of planets in the disk. We check the possibility of this dependence by calculating the accuracy for each model for disks with a given number of planets. Figure~\ref{fig:acc_N} gives the results of these calculations. The $p$-values from Pearson correlation tests shown in Figure~\ref{fig:p_values} demonstrate that only the 47-channel RegNet using a decision threshold of 50$\%$ has a statistically significant relationship between accuracy and the number of planets ($p < 0.05$). While the 47-channel RegNet using a decision threshold of 95$\%$ does not have a statistically significant relationship, it is clear that the performance is worse when there are planets, i.e. there are many false negatives. This is in line with this model's relatively low AUC and Figure~\ref{fig:conf_mat}. The performance of all other models does not differ significantly with the number of planets.

\subsection{Simulation Activations} \label{ssec:sim_activations}

\begin{figure*}
    \centering
    \includegraphics[width=0.95\linewidth]{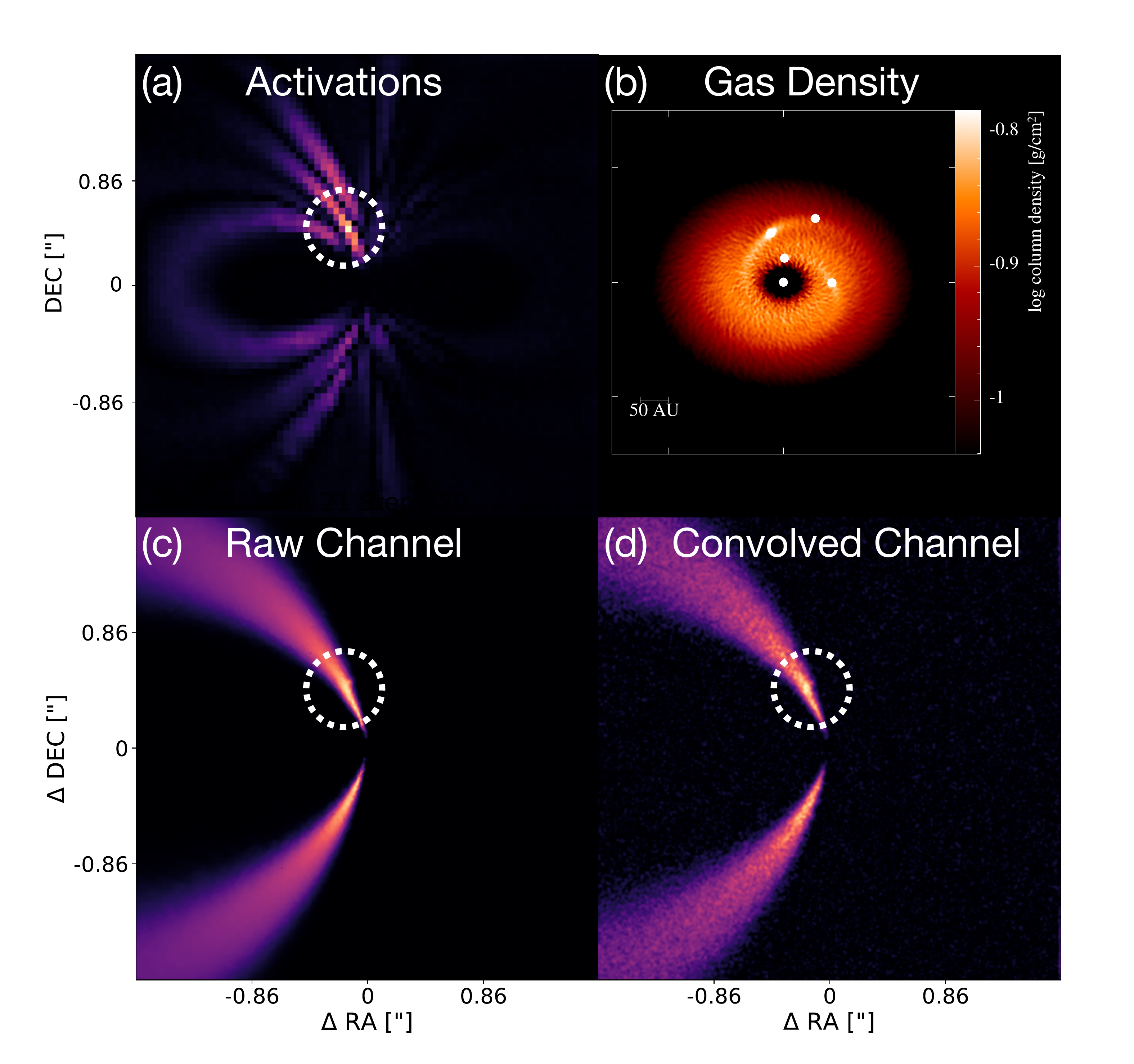}
    \caption{(a): Mean-subtracted activations using the 47-channel EfficientNetV2. The innermost planet has M=0.28M$_{\rm{J}}$, the second planet has M=0.20M$_{\rm{J}}$, the third planet has M=0.36M$_{\rm{J}}$, and the outermost planet has has M=0.02M$_{\rm{J}}$. (b): Disk column density after an azimuthal rotation of $\phi=131^{\circ}$ and inclined by $i=36^{\circ}$. (c): Unconvolved velocity channel with the most obvious kink. (d) Corresponding convolved velocity channel. Based on the kink, the approximate location of the planet causing the most significant kink is in the white circle. The planets were correctly predicted with a confidence of $>99\%$.}
    \label{fig:run_21}
\end{figure*}



\par 

Many disks do not show obvious signs of planets, either in continuum or line emission. Inspecting the activations of the disks helps overcome this. Figure~\ref{fig:run_21} shows the result from a disk that is difficult to classify by eye. The models predict the presence of an exoplanet with high confidence, activating on a subset of channels that are deemed important by the model. 
This gives information on the azimuthal location, and the radial location may be inferred by either the location with the strongest kink or coincident gaps in continuum images. 

%
\par
Panels (c) and (d) in Figure~\ref{fig:run_21} show the raw and convolved velocity channels, respectively. If one were simply inspecting the velocity channels by eye, it would be trivially easy to miss the kink even in the unconvolved data. The convolved, noisy data shows essentially no sign of the planets. Despite lacking any clear signatures that can be observed visually, the models correctly classify the disks and locate the planets. This demonstrates that our models are able to accurately find planets that the by-eye method may fail to find.

\begin{figure}
    \centering
    \includegraphics[width=1.0\columnwidth]{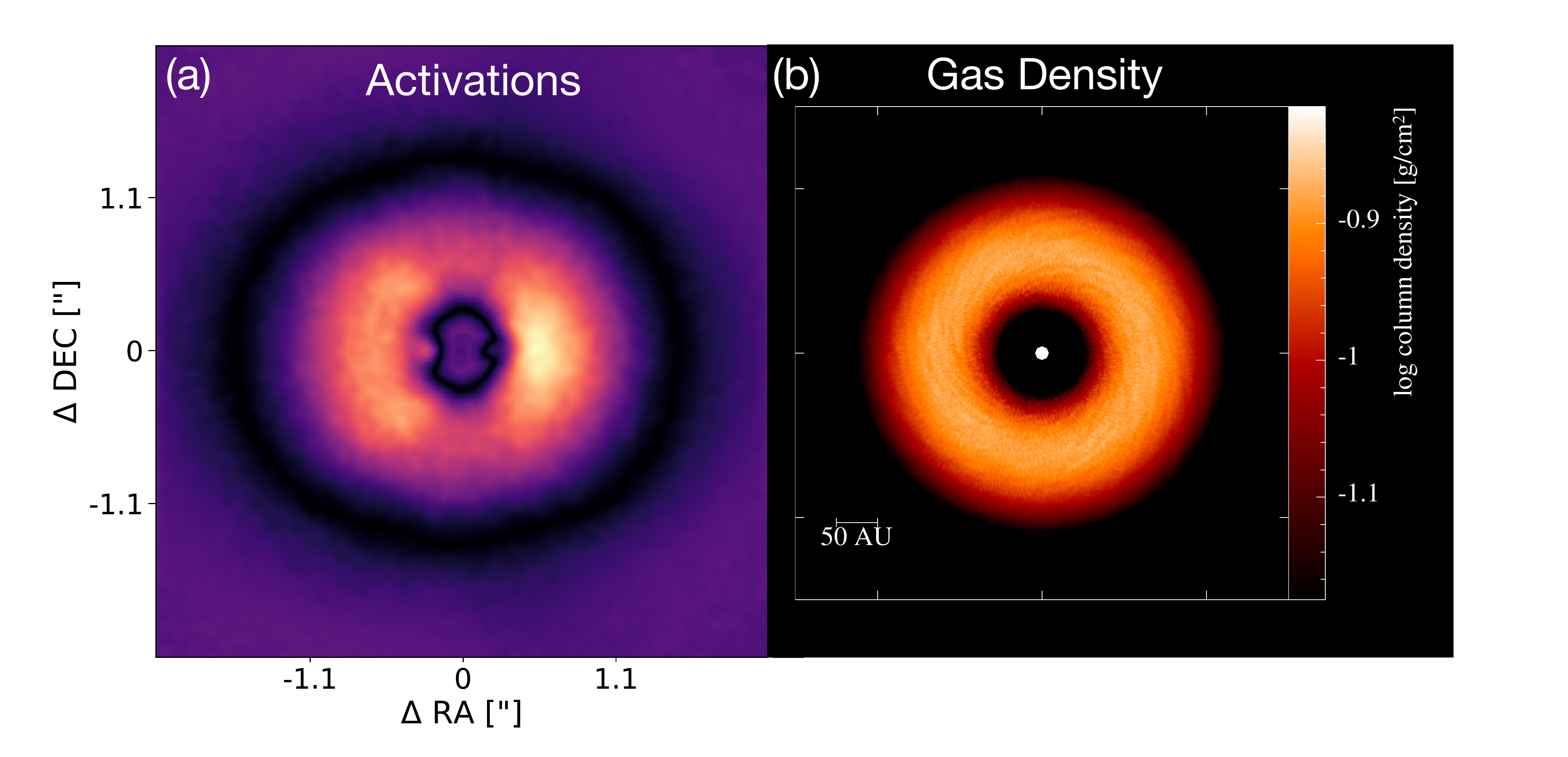}
    \caption{(a): Typical example of mean-subtracted activations for a disk with no planet. (b): Disk column density after an azimuthal rotation of $\phi=153^{\circ}$ and inclined by $i=13^{\circ}$.}
    \label{fig:no_planet}
\end{figure}

\par 
Figure~\ref{fig:no_planet} shows a typical activation for a disk with no planet. 
The entire disk is activated equally, indicating that there is no localized region that exhibits behaviors indicative of a planet.

\subsection{Application to Real Observations} \label{ssec:obs_activations}

\begin{figure*}
    \centering
    \includegraphics[width=0.95\linewidth]{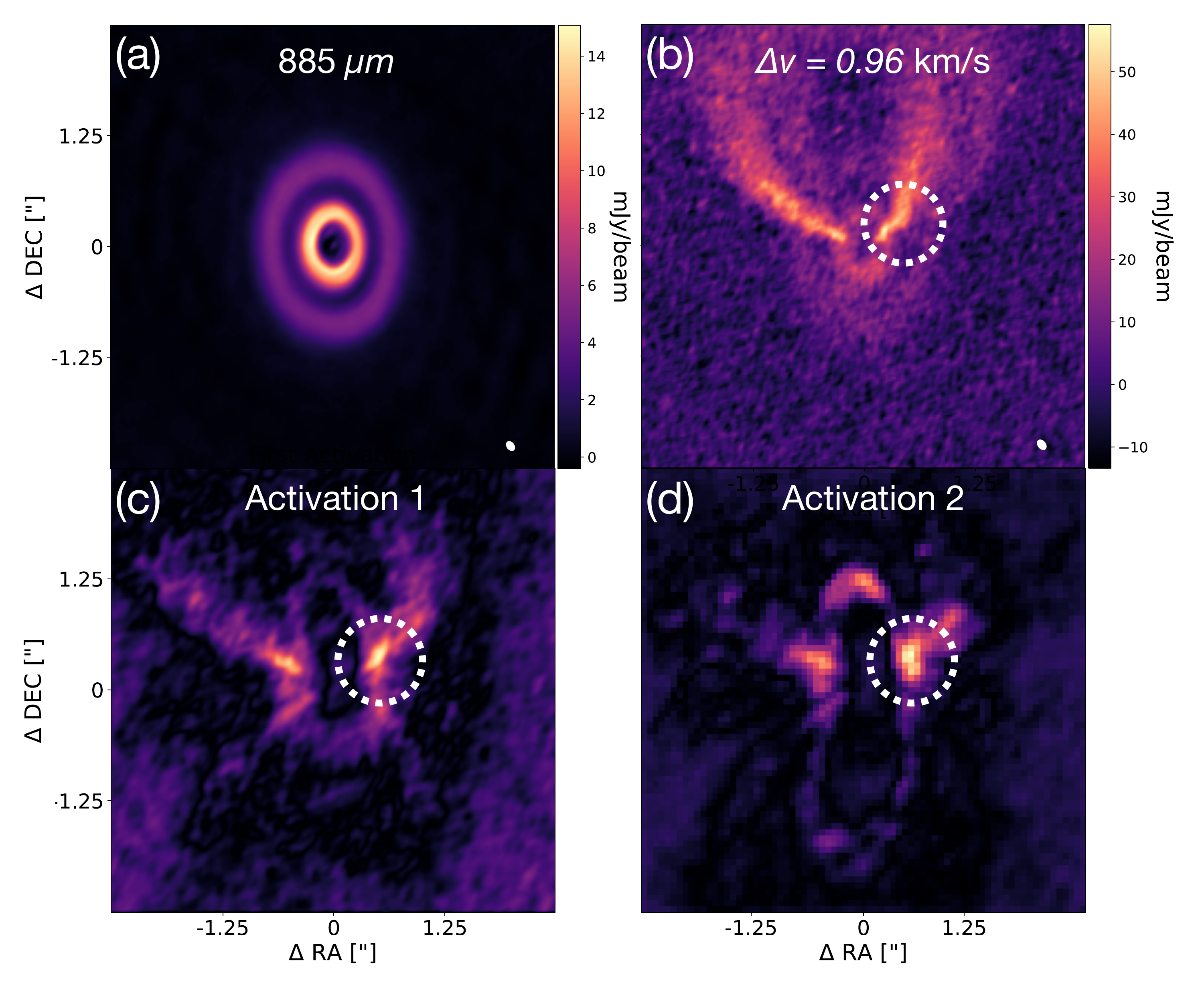}
    \caption{(a): HD 97048 at 885 $\mu m$. (b): $\Delta v = 0.96$ km/s channel. Bottom: Mean-subtracted activations from two different layers for HD 97048 using the 47-channel RegNet. The estimated planet location is given inside the white circle. The beams for the observations are given in the bottom right corner.}
    \label{fig:hd_97048}
\end{figure*}

Next, we demonstrate the effectiveness of our model by applying it to real telescope observations. Kinematic detections of planets have been demonstrated for many disks~\citep{Pinte2020}, but here we focus on the HD 163296 and HD 97048 disks as a proof of concept~\citep{Pinte2018, Pinte2019}. 
Using the same data as~\citet{Pinte2018} and~\citet{Pinte2019}, we demonstrate that our models replicate the prediction and estimated location of the forming exoplanet. Figure~\ref{fig:hd_97048} (a) shows the HD 97048 in continuum, and panel (b) shows the velocity channel with the planet's signature circled.

\begin{table*}

    \centering
    \begin{tabular}{lccc}

    \hline 
    Model       &     HD 97048 Softmax   & HD 163296 Softmax\\
    \hline 
    EffecientNetV2: 47 Channels  & $85\%$    & $37\%$    \\
    EffecientNetV2: 61 Channels  & $89\%$    & $69\%$    \\
    EffecientNetV2: 75 Channels  & $64\%$    & $73\%$    \\
    RegNet: 47 Channels          & $>99\%$   & $74\%$    \\
    RegNet: 61 Channels          & $96\%$    & $>99\%$   \\
    RegNet: 75 Channels          & $>99\%$   & $>99\%$   \\
    \hline
    \end{tabular}
    
\caption{Prediction confidences for all models tested on HD 97048 and HD 163296.}
\label{tab:obs_results}
\end{table*}

\par 
Table~\ref{tab:obs_results} shows the softmax values for the observations. For HD 97048, all RegNet models correctly predict a planet with $> 95\%$ confidence. Two EfficientNetV2 models correctly predict a planet with $\geq 85$\% confidence. 
Figure~\ref{fig:hd_97048} shows that the velocity channels that contain the planet are strongly activated. The activation in Figure~\ref{fig:hd_97048} (d) is strong on the planet itself rather than simply the entire velocity channel. 

The results for HD 163296 are not as confident,  
but the two best-performing models predict the presence of at least one planet with over 99\% confidence. 

Both classes of model perform better when there are more input channels, a trend not seen for HD 97048. A possible explanation is the fact that there are about twice as many velocity channels that cover HD 163296 when compared to HD 97048. While HD 97048 has roughly 75 channels covering the disk, HD 163296 has roughly 150. All models applied to HD 163296 must therefore combine significantly more velocity channels, which may decrease its ability to extract information from the images. The activation structure tends to be less informative, perhaps for the similar reasons, so it is omitted. Despite this limitation, the high confidence of some of the models should encourage a human observer to seriously scrutinize the kinematic structure.

\subsection{Limitations and Future Work} \label{ssec:limitations}

Our work comes with several limitations. All simulations were run with SPH, so we did not demonstrate that our methods would also be applicable to data from grid-based codes. However, there is typically a qualitative agreement between results from SPH and grid-based methods~\citep{val_borro_2006, agertz_2007}. A good direction for future work would be to train on simulation data from a variety of hydrodynamics codes, and check that the same results can be obtained.


Any differences in performance of the models between the codes, e.g. if the models consistently performed better on SPH data, would be difficult to interpret and the cause could be hard to identify. It is possible that the method of domain adaptation~\citep{domain_adaptation}, which has already found success in astronomy~\citep{Vilalta_2019, Alexander_2021, Cip_2022}, could be used to encourage the models to overcome any differences between datasets. This is an avenue that is ripe for exploration in future work. 


In order to keep the parameter space manageable and demonstrate a proof of concept in this work, we therefore trained on data produced by only one hydrodynamics code (\texttt{PHANTOM}).


\par
The absence of true image segmentation~\citep{segmentation} or object detection~\citep{object_detection} during training is another limitation that requires addressing in future work. While we present results in the form of activation structures that suggest the location(s) of the planet(s), these activations do not always perfectly coincide with the planets, nor capture all planets. User analysis is required in order to pinpoint the radial (and sometimes azimuthal) location(s) of the planets even with the information given by the activation structures. The main utility of the presented models is therefore to give a user an indication as to whether a planet is present and motivate and direct the analysis of the velocity channels. Segmentation and object detection would address these shortcomings by precisely pinpointing the location(s) of the planet(s) without any necessary further analysis.
\par

\par
The assertion of a vertically isothermal disk is a simplifying assumption that is commonly used in simulations of protoplantary discs \citep[see, e.g.,][]{dipierro2015,Pinte2019} . It has been demonstrated that the dispersion relation and morphology of the planet's signature can change significantly if the disk is vertically stratified~\citep{Ogilvie2002, Juhasz2018}. 

However, the presence of any disturbance is more important than its specific morphology. There does exist the possibility that vertical stratification dampens the signatures to an extent that the models' performances appreciably decrease, but our results demonstrate that even extremely subtle signals can be found by the models. Regardless, a more realistic treatment of the thermal structure may be useful in future works.

\section{Conclusions} \label{sec:conclusion}
Our results show that a machine learning model trained on synthetic data can determine the presence and location of forming exoplanets in real telescope observations of protoplanetary accretion disks. We apply this model to the HD 97048 and HD 163296 systems. Multiple models predict the presence of a planet with greater than $99\%$ confidence in both systems. The presence and location of the planet in HD 97048 is corroborated by activating strongly on the same location given by~\citet{Pinte2019}. 
\par 
Five out of six models have greater than 90\% accuracy when using a 50\% decision threshold. When using a 95\% decision threshold, four out of six models have an accuracy greater than 90\%. Five models have an AUC of at least 0.98. While EfficientNetV2 and RegNet perform comparably on the synthetic test data, RegNet outperforms EfficientNetV2 on the real telescope observations.

\par 
This work is the first step in a series of works that use machine learning to automate the detection and analysis of objects within protoplanetary disks. Applying full segmentation and object detection methods will further increase the ability of our models to locate planets and infer properties, such as planetary mass. We leave this work to the future. We envision final models that can be used easily for multiple purposes and be distributed freely throughout the community. 

\par
As ALMA continues to deliver larger and larger disk survey datasets and next generation telescopes \textemdash such as JWST, ngVLA  and the SKA \textemdash come online, the time-cost of analyzing more and more data will increase. Our proof-of-concept work demonstrates that even when high confidence is demanded, machine learning can automate this task and match and exceed human proficiency.

\section{Acknowledgements}

We would like to thank the referee for insightful comments that improved the paper. This paper makes use of the following ALMA data: ADS/JAO.ALMA\#2016.1.00826.S and ADS/JAO.ALMA\#2013.1.00601.S. ALMA is a partnership of ESO (representing its member states), NSF (USA) and NINS (Japan), together with NRC (Canada), MOST and ASIAA (Taiwan), and KASI (Republic of Korea), in cooperation with the Republic of Chile. The Joint ALMA Observatory is operated by ESO, AUI/NRAO and NAOJ. The National Radio Astronomy Observatory is a facility of the National Science Foundation operated under cooperative agreement by Associated Universities, Inc. SPH results are visualized using \texttt{SPLASH}~\citep{splash}. JT was a participant in the 2022 Google Summer of Code program. This study was supported in part by resources and technical expertise from the Georgia Advanced Computing Resource Center, a partnership between the University of Georgia’s Office of the Vice President for Research and Office of the Vice President for Information Technology.

\bibliography{paper}{}

\begin{thebibliography}{}
\expandafter\ifx\csname natexlab\endcsname\relax\def\natexlab#1{#1}\fi
\providecommand{\url}[1]{\href{#1}{#1}}
\providecommand{\dodoi}[1]{doi:~\href{http://doi.org/#1}{\nolinkurl{#1}}}
\providecommand{\doeprint}[1]{\href{http://ascl.net/#1}{\nolinkurl{http://ascl.net/#1}}}
\providecommand{\doarXiv}[1]{\href{https://arxiv.org/abs/#1}{\nolinkurl{https://arxiv.org/abs/#1}}}

\bibitem[{{Agertz} {et~al.}(2007){Agertz}, {Moore}, {Stadel}, {Potter},
  {Miniati}, {Read}, {Mayer}, {Gawryszczak}, {Kravtsov}, {Nordlund}, {Pearce},
  {Quilis}, {Rudd}, {Springel}, {Stone}, {Tasker}, {Teyssier}, {Wadsley}, \&
  {Walder}}]{agertz_2007}
{Agertz}, O., {Moore}, B., {Stadel}, J., {et~al.} 2007, \mnras, 380, 963,
  \dodoi{10.1111/j.1365-2966.2007.12183.x}

\bibitem[{{Alexander} {et~al.}(2020){Alexander}, {Gleyzer}, {McDonough},
  {Toomey}, \& {Usai}}]{alexander_2020}
{Alexander}, S., {Gleyzer}, S., {McDonough}, E., {Toomey}, M.~W., \& {Usai}, E.
  2020, \apj, 893, 15, \dodoi{10.3847/1538-4357/ab7925}

\bibitem[{{Alexander} {et~al.}(2021){Alexander}, {Gleyzer}, {Reddy}, {Tidball},
  \& {Toomey}}]{Alexander_2021}
{Alexander}, S., {Gleyzer}, S., {Reddy}, P., {Tidball}, M., \& {Toomey}, M.~W.
  2021, arXiv e-prints, arXiv:2112.12121.
\newblock \doarXiv{2112.12121}

\bibitem[{Andrews {et~al.}(2009)Andrews, Wilner, Hughes, Qi, \&
  Dullemond}]{Andrews2009}
Andrews, S.~M., Wilner, D.~J., Hughes, A.~M., Qi, C., \& Dullemond, C.~P. 2009,
  The Astrophysical Journal, 700, 1502, \dodoi{10.1088/0004-637x/700/2/1502}

\bibitem[{{Andrews} {et~al.}(2018){Andrews}, {Huang}, {P{\'e}rez}, {Isella},
  {Dullemond}, {Kurtovic}, {Guzm{\'a}n}, {Carpenter}, {Wilner}, {Zhang}, {Zhu},
  {Birnstiel}, {Bai}, {Benisty}, {Hughes}, {{\"O}berg}, \& {Ricci}}]{DSHARP}
{Andrews}, S.~M., {Huang}, J., {P{\'e}rez}, L.~M., {et~al.} 2018, \apjl, 869,
  L41, \dodoi{10.3847/2041-8213/aaf741}

\bibitem[{{Auddy} {et~al.}(2022){Auddy}, {Dey}, {Lin}, {Carrera}, \&
  {Simon}}]{Auddy2022}
{Auddy}, S., {Dey}, R., {Lin}, M.-K., {Carrera}, D., \& {Simon}, J.~B. 2022,
  \apj, 936, 93, \dodoi{10.3847/1538-4357/ac7a3c}

\bibitem[{{Auddy} {et~al.}(2021){Auddy}, {Dey}, {Lin}, \& {Hall}}]{Auddy2021}
{Auddy}, S., {Dey}, R., {Lin}, M.-K., \& {Hall}, C. 2021, \apj, 920, 3,
  \dodoi{10.3847/1538-4357/ac1518}

\bibitem[{{Barraza-Alfaro} {et~al.}(2021){Barraza-Alfaro}, {Flock}, {Marino},
  \& {P{\'e}rez}}]{Barraza2021}
{Barraza-Alfaro}, M., {Flock}, M., {Marino}, S., \& {P{\'e}rez}, S. 2021, \aap,
  653, A113, \dodoi{10.1051/0004-6361/202140535}

\bibitem[{Ben-David {et~al.}(2010)Ben-David, Blitzer, Crammer, Kulesza,
  Pereira, \& Vaughan}]{domain_adaptation}
Ben-David, S., Blitzer, J., Crammer, K., {et~al.} 2010, Machine Learning, 79,
  151, \dodoi{10.1007/s10994-009-5152-4}

\bibitem[{Biewald(2020)}]{wandb}
Biewald, L. 2020, Experiment Tracking with Weights and Biases.
\newblock \url{https://www.wandb.com/}

\bibitem[{{{\'C}iprijanovi{\'c}} {et~al.}(2022){{\'C}iprijanovi{\'c}},
  {Kafkes}, {Snyder}, {S{\'a}nchez}, {Perdue}, {Pedro}, {Nord}, {Madireddy}, \&
  {Wild}}]{Cip_2022}
{{\'C}iprijanovi{\'c}}, A., {Kafkes}, D., {Snyder}, G., {et~al.} 2022, Machine
  Learning: Science and Technology, 3, 035007, \dodoi{10.1088/2632-2153/ac7f1a}

\bibitem[{{Cire{\c{s}}an} {et~al.}(2011){Cire{\c{s}}an}, {Meier}, {Masci},
  {Gambardella}, \& {Schmidhuber}}]{mnist}
{Cire{\c{s}}an}, D.~C., {Meier}, U., {Masci}, J., {Gambardella}, L.~M., \&
  {Schmidhuber}, J. 2011, arXiv e-prints, arXiv:1102.0183.
\newblock \doarXiv{1102.0183}

\bibitem[{{de Val-Borro} {et~al.}(2006){de Val-Borro}, {Edgar}, {Artymowicz},
  {Ciecielag}, {Cresswell}, {D'Angelo}, {Delgado-Donate}, {Dirksen}, {Fromang},
  {Gawryszczak}, {Klahr}, {Kley}, {Lyra}, {Masset}, {Mellema}, {Nelson},
  {Paardekooper}, {Peplinski}, {Pierens}, {Plewa}, {Rice}, {Sch{\"a}fer}, \&
  {Speith}}]{val_borro_2006}
{de Val-Borro}, M., {Edgar}, R.~G., {Artymowicz}, P., {et~al.} 2006, \mnras,
  370, 529, \dodoi{10.1111/j.1365-2966.2006.10488.x}

\bibitem[{{Dipierro} {et~al.}(2015){Dipierro}, {Price}, {Laibe}, {Hirsh},
  {Cerioli}, \& {Lodato}}]{dipierro2015}
{Dipierro}, G., {Price}, D., {Laibe}, G., {et~al.} 2015, \mnras, 453, L73,
  \dodoi{10.1093/mnrasl/slv105}

\bibitem[{{Dosovitskiy} {et~al.}(2020){Dosovitskiy}, {Beyer}, {Kolesnikov},
  {Weissenborn}, {Zhai}, {Unterthiner}, {Dehghani}, {Minderer}, {Heigold},
  {Gelly}, {Uszkoreit}, \& {Houlsby}}]{Dosovitskiy_2020}
{Dosovitskiy}, A., {Beyer}, L., {Kolesnikov}, A., {et~al.} 2020, arXiv
  e-prints, arXiv:2010.11929.
\newblock \doarXiv{2010.11929}

\bibitem[{{Draine} \& {Lee}(1984)}]{Draine1984}
{Draine}, B.~T., \& {Lee}, H.~M. 1984, \apj, 285, 89, \dodoi{10.1086/162480}

\bibitem[{{Elfwing} {et~al.}(2017){Elfwing}, {Uchibe}, \& {Doya}}]{silu}
{Elfwing}, S., {Uchibe}, E., \& {Doya}, K. 2017, arXiv e-prints,
  arXiv:1702.03118.
\newblock \doarXiv{1702.03118}

\bibitem[{Glorot {et~al.}(2011)Glorot, Bordes, \& Bengio}]{relu}
Glorot, X., Bordes, A., \& Bengio, Y. 2011, in AISTATS

\bibitem[{Goodfellow {et~al.}(2016)Goodfellow, Bengio, \& Courville}]{softmax}
Goodfellow, I., Bengio, Y., \& Courville, A. 2016, Deep Learning (MIT Press)

\bibitem[{{Hall} {et~al.}(2020){Hall}, {Dong}, {Teague}, {Terry}, {Pinte},
  {Paneque-Carre{\~n}o}, {Veronesi}, {Alexander}, \& {Lodato}}]{hall2020}
{Hall}, C., {Dong}, R., {Teague}, R., {et~al.} 2020, \apj, 904, 148,
  \dodoi{10.3847/1538-4357/abac17}

\bibitem[{Hanley \& McNeil(1982)}]{auc}
Hanley, J.~A., \& McNeil, B.~J. 1982, Radiology, 143 1, 29

\bibitem[{{He} {et~al.}(2015){He}, {Zhang}, {Ren}, \& {Sun}}]{resnet}
{He}, K., {Zhang}, X., {Ren}, S., \& {Sun}, J. 2015, arXiv e-prints,
  arXiv:1512.03385.
\newblock \doarXiv{1512.03385}

\bibitem[{Hinton {et~al.}(1995)Hinton, Dayan, Frey, \& Neal}]{cross_entropy}
Hinton, G.~E., Dayan, P., Frey, B.~J., \& Neal, R.~M. 1995, Science, 268, 1158.
\newblock \url{http://www.jstor.org/stable/2888376}

\bibitem[{{Huang} {et~al.}(2018){Huang}, {Andrews}, {Dullemond}, {Isella},
  {P{\'e}rez}, {Guzm{\'a}n}, {{\"O}berg}, {Zhu}, {Zhang}, {Bai}, {Benisty},
  {Birnstiel}, {Carpenter}, {Hughes}, {Ricci}, {Weaver}, \&
  {Wilner}}]{DSHARP_2}
{Huang}, J., {Andrews}, S.~M., {Dullemond}, C.~P., {et~al.} 2018, \apjl, 869,
  L42, \dodoi{10.3847/2041-8213/aaf740}

\bibitem[{{Jo} \& {Kim}(2019)}]{jo2019}
{Jo}, Y., \& {Kim}, J.-h. 2019, \mnras, 489, 3565,
  \dodoi{10.1093/mnras/stz2304}

\bibitem[{{Juh{\'a}sz} \& {Rosotti}(2018)}]{Juhasz2018}
{Juh{\'a}sz}, A., \& {Rosotti}, G.~P. 2018, \mnras, 474, L32,
  \dodoi{10.1093/mnrasl/slx182}

\bibitem[{{Kingma} \& {Ba}(2014)}]{adam}
{Kingma}, D.~P., \& {Ba}, J. 2014, arXiv e-prints, arXiv:1412.6980.
\newblock \doarXiv{1412.6980}

\bibitem[{McKay {et~al.}(1979)McKay, Beckman, \& Conover}]{LHC}
McKay, M.~D., Beckman, R.~J., \& Conover, W.~J. 1979, Technometrics, 21, 239.
\newblock \url{http://www.jstor.org/stable/1268522}

\bibitem[{{Minaee} {et~al.}(2020){Minaee}, {Boykov}, {Porikli}, {Plaza},
  {Kehtarnavaz}, \& {Terzopoulos}}]{segmentation}
{Minaee}, S., {Boykov}, Y., {Porikli}, F., {et~al.} 2020, arXiv e-prints,
  arXiv:2001.05566.
\newblock \doarXiv{2001.05566}

\bibitem[{{M{\"o}ller} \& {de Boissi{\`e}re}(2020)}]{moller20}
{M{\"o}ller}, A., \& {de Boissi{\`e}re}, T. 2020, \mnras, 491, 4277,
  \dodoi{10.1093/mnras/stz3312}

\bibitem[{{{\"O}berg} {et~al.}(2021){{\"O}berg}, {Guzm{\'a}n}, {Walsh},
  {Aikawa}, {Bergin}, {Law}, {Loomis}, {Alarc{\'o}n}, {Andrews}, {Bae},
  {Bergner}, {Boehler}, {Booth}, {Bosman}, {Calahan}, {Cataldi}, {Cleeves},
  {Czekala}, {Furuya}, {Huang}, {Ilee}, {Kurtovic}, {Le Gal}, {Liu}, {Long},
  {M{\'e}nard}, {Nomura}, {P{\'e}rez}, {Qi}, {Schwarz}, {Sierra}, {Teague},
  {Tsukagoshi}, {Yamato}, {van't Hoff}, {Waggoner}, {Wilner}, \&
  {Zhang}}]{MAPS}
{{\"O}berg}, K.~I., {Guzm{\'a}n}, V.~V., {Walsh}, C., {et~al.} 2021, \apjs,
  257, 1, \dodoi{10.3847/1538-4365/ac1432}

\bibitem[{{Ogilvie} \& {Lubow}(2002)}]{Ogilvie2002}
{Ogilvie}, G.~I., \& {Lubow}, S.~H. 2002, \mnras, 330, 950,
  \dodoi{10.1046/j.1365-8711.2002.05148.x}

\bibitem[{{Paszke} {et~al.}(2019){Paszke}, {Gross}, {Massa}, {Lerer},
  {Bradbury}, {Chanan}, {Killeen}, {Lin}, {Gimelshein}, {Antiga}, {Desmaison},
  {K{\"o}pf}, {Yang}, {DeVito}, {Raison}, {Tejani}, {Chilamkurthy}, {Steiner},
  {Fang}, {Bai}, \& {Chintala}}]{pytorch}
{Paszke}, A., {Gross}, S., {Massa}, F., {et~al.} 2019, arXiv e-prints,
  arXiv:1912.01703.
\newblock \doarXiv{1912.01703}

\bibitem[{{Pinte} {et~al.}(2009){Pinte}, {Harries}, {Min}, {Watson},
  {Dullemond}, {Woitke}, {M{\'e}nard}, \& {Dur{\'a}n-Rojas}}]{Pinte2009}
{Pinte}, C., {Harries}, T.~J., {Min}, M., {et~al.} 2009, \aap, 498, 967,
  \dodoi{10.1051/0004-6361/200811555}

\bibitem[{{Pinte} {et~al.}(2006){Pinte}, {M{\'e}nard}, {Duch{\^e}ne}, \&
  {Bastien}}]{Pinte2006}
{Pinte}, C., {M{\'e}nard}, F., {Duch{\^e}ne}, G., \& {Bastien}, P. 2006, \aap,
  459, 797, \dodoi{10.1051/0004-6361:20053275}

\bibitem[{{Pinte} {et~al.}(2018){Pinte}, {Price}, {M{\'e}nard}, {Duch{\^e}ne},
  {Dent}, {Hill}, {de Gregorio-Monsalvo}, {Hales}, \& {Mentiplay}}]{Pinte2018}
{Pinte}, C., {Price}, D.~J., {M{\'e}nard}, F., {et~al.} 2018, \apjl, 860, L13,
  \dodoi{10.3847/2041-8213/aac6dc}

\bibitem[{{Pinte} {et~al.}(2019){Pinte}, {van der Plas}, {M{\'e}nard}, {Price},
  {Christiaens}, {Hill}, {Mentiplay}, {Ginski}, {Choquet}, {Boehler},
  {Duch{\^e}ne}, {Perez}, \& {Casassus}}]{Pinte2019}
{Pinte}, C., {van der Plas}, G., {M{\'e}nard}, F., {et~al.} 2019, Nature
  Astronomy, 3, 1109, \dodoi{10.1038/s41550-019-0852-6}

\bibitem[{{Pinte} {et~al.}(2020){Pinte}, {Price}, {M{\'e}nard}, {Duch{\^e}ne},
  {Christiaens}, {Andrews}, {Huang}, {Hill}, {van der Plas}, {Perez}, {Isella},
  {Boehler}, {Dent}, {Mentiplay}, \& {Loomis}}]{Pinte2020}
{Pinte}, C., {Price}, D.~J., {M{\'e}nard}, F., {et~al.} 2020, \apjl, 890, L9,
  \dodoi{10.3847/2041-8213/ab6dda}

\bibitem[{{Price}(2007)}]{splash}
{Price}, D.~J. 2007, \pasa, 24, 159, \dodoi{10.1071/AS07022}

\bibitem[{{Price} {et~al.}(2018){Price}, {Wurster}, {Tricco}, {Nixon},
  {Toupin}, {Pettitt}, {Chan}, {Mentiplay}, {Laibe}, {Glover}, {Dobbs},
  {Nealon}, {Liptai}, {Worpel}, {Bonnerot}, {Dipierro}, {Ballabio}, {Ragusa},
  {Federrath}, {Iaconi}, {Reichardt}, {Forgan}, {Hutchison}, {Constantino},
  {Ayliffe}, {Hirsh}, \& {Lodato}}]{phantom18}
{Price}, D.~J., {Wurster}, J., {Tricco}, T.~S., {et~al.} 2018, \pasa, 35, e031,
  \dodoi{10.1017/pasa.2018.25}

\bibitem[{{Siess} {et~al.}(2000){Siess}, {Dufour}, \& {Forestini}}]{siess}
{Siess}, L., {Dufour}, E., \& {Forestini}, M. 2000, \aap, 358, 593.
\newblock \doarXiv{astro-ph/0003477}

\bibitem[{{Simon} {et~al.}(2015){Simon}, {Hughes}, {Flaherty}, {Bai}, \&
  {Armitage}}]{Simon_2015}
{Simon}, J.~B., {Hughes}, A.~M., {Flaherty}, K.~M., {Bai}, X.-N., \&
  {Armitage}, P.~J. 2015, \apj, 808, 180, \dodoi{10.1088/0004-637X/808/2/180}

\bibitem[{{Szegedy} {et~al.}(2014){Szegedy}, {Liu}, {Jia}, {Sermanet}, {Reed},
  {Anguelov}, {Erhan}, {Vanhoucke}, \& {Rabinovich}}]{inception}
{Szegedy}, C., {Liu}, W., {Jia}, Y., {et~al.} 2014, arXiv e-prints,
  arXiv:1409.4842.
\newblock \doarXiv{1409.4842}

\bibitem[{{Tan} \& {Le}(2019)}]{effnet}
{Tan}, M., \& {Le}, Q.~V. 2019, arXiv e-prints, arXiv:1905.11946.
\newblock \doarXiv{1905.11946}

\bibitem[{{Tan} \& {Le}(2021)}]{effnet_v2}
---. 2021, arXiv e-prints, arXiv:2104.00298.
\newblock \doarXiv{2104.00298}

\bibitem[{{Teague} {et~al.}(2018){Teague}, {Bae}, {Bergin}, {Birnstiel}, \&
  {Foreman-Mackey}}]{Teague2018}
{Teague}, R., {Bae}, J., {Bergin}, E.~A., {Birnstiel}, T., \& {Foreman-Mackey},
  D. 2018, \apjl, 860, L12, \dodoi{10.3847/2041-8213/aac6d7}

\bibitem[{{Terry} {et~al.}(2022){Terry}, {Hall}, {Longarini}, {Lodato}, {Toci},
  {Veronesi}, {Paneque-Carre{\~n}o}, \& {Pinte}}]{terry2022}
{Terry}, J.~P., {Hall}, C., {Longarini}, C., {et~al.} 2022, \mnras, 510, 1671,
  \dodoi{10.1093/mnras/stab3513}

\bibitem[{{Vaswani} {et~al.}(2017){Vaswani}, {Shazeer}, {Parmar}, {Uszkoreit},
  {Jones}, {Gomez}, {Kaiser}, \& {Polosukhin}}]{Vaswani_2017}
{Vaswani}, A., {Shazeer}, N., {Parmar}, N., {et~al.} 2017, arXiv e-prints,
  arXiv:1706.03762.
\newblock \doarXiv{1706.03762}

\bibitem[{{Vilalta} {et~al.}(2019){Vilalta}, {Dhar Gupta}, {Boumber}, \&
  {Meskhi}}]{Vilalta_2019}
{Vilalta}, R., {Dhar Gupta}, K., {Boumber}, D., \& {Meskhi}, M.~M. 2019, \pasp,
  131, 108008, \dodoi{10.1088/1538-3873/aaf1fc}

\bibitem[{{Xu} {et~al.}(2021){Xu}, {Pan}, {Pan}, {Hoi}, {Yi}, \& {Xu}}]{regnet}
{Xu}, J., {Pan}, Y., {Pan}, X., {et~al.} 2021, arXiv e-prints,
  arXiv:2101.00590.
\newblock \doarXiv{2101.00590}

\bibitem[{{Zhao} {et~al.}(2018){Zhao}, {Zheng}, {Xu}, \&
  {Wu}}]{object_detection}
{Zhao}, Z.-Q., {Zheng}, P., {Xu}, S.-t., \& {Wu}, X. 2018, arXiv e-prints,
  arXiv:1807.05511.
\newblock \doarXiv{1807.05511}

\bibitem[{{Zhou} {et~al.}(2021){Zhou}, {Yang}, {Yu}, {Liu}, {Duan}, {Weng},
  {Chen}, {Liang}, {Fang}, {Zhou}, {Ju}, {Luo}, {Guo}, {Ma}, {Xie}, {Wang}, \&
  {Zhou}}]{Zhou2021}
{Zhou}, W., {Yang}, Y., {Yu}, C., {et~al.} 2021, Nature Communications, 12,
  1259, \dodoi{10.1038/s41467-021-21466-z}

\end{thebibliography}
\bibliographystyle{aasjournal}



\end{document}